\documentclass[aps,pra,twocolumn,superscriptaddress,notitlepage,nofootinbib,longbibliography]{revtex4-2}
\usepackage{adjustbox}
\usepackage{lipsum} 
\usepackage{mathrsfs}
\usepackage{epsfig}
\usepackage{graphicx}
\usepackage{amsfonts}
\usepackage[figuresright]{rotating}
\usepackage{amssymb}
\usepackage{amsmath}
\usepackage{dcolumn}
\usepackage{bm}
\usepackage{dutchcal}
\usepackage{xcolor}
\usepackage{color}
\usepackage{braket}
\usepackage{units}
\usepackage{xspace}

\usepackage{bbold}
\definecolor{mypink3}{cmyk}{0, 0.7808, 0.4429, 0.1412}
\definecolor{mypink1}{rgb}{0.858, 0.188, 0.478}
\definecolor{mypink2}{RGB}{219, 48, 122}
\usepackage[colorlinks=true, allcolors=mypink1]{hyperref}
\usepackage[fleqn]{mathtools}

\usepackage[shortlabels]{enumitem}

\newcommand{\ECNU}{Quantum Institute for Light and Atoms, School of Physics, East China Normal University, Shanghai 200062, China}
\newcommand{\HB}{Shanghai Branch, Hefei National Laboratory, Shanghai 201315, China}
\newcommand{\SPA}{School of Physics and Astronomy, and Tsung-Dao Lee Institute, Shanghai Jiao Tong University, Shanghai 200240, China}
\newcommand{\SR}{Shanghai Research Center for Quantum Sciences, Shanghai 201315, China}
\newcommand{\CIC}{Collaborative Innovation Center of Extreme Optics, Shanxi University, Taiyuan, Shanxi 030006, China}
\newcommand{\NTU}{School of Physical Science and Technology, Nantong University, Nantong 226019, People’s Republic of China}   

\begin{document}
\title{Quantum-Chaos-Assisted Measure Synchronization in a Double-Well Ultracold Bose Gas}

\author{Xianchao Le}
\affiliation{\ECNU}
\affiliation{\HB}
\author{Jiaxin Peng}  
\affiliation{\NTU}
\author{Weiping Zhang}
\affiliation{\HB}
\affiliation{\SPA}
\affiliation{\SR}
\affiliation{\CIC}
\author{Keye Zhang}
\email{kyzhang@phy.ecnu.edu.cn}
\affiliation{\ECNU}
\affiliation{\HB}
\date{\today}

\begin{abstract}
This paper investigates measure synchronization (MS) in a two-component ultracold atomic system confined in a double-well potential, which realizes a quantum many-body bosonic Josephson junction. Through spectral-statistics analysis, we identify the quantum chaotic parameter regime and confirm that the system achieves MS within this regime with weak dependence on the initial energy imbalance. We further show that chaotic dynamics not only significantly shortens the time scale for establishing synchronization but is also accompanied by entanglement entropy approaching the random-state prediction. These results reveal close connections among quantum chaos, MS, and quantum entanglement in conservative quantum many-body systems and may motivate further studies on the underlying mechanisms of chaos-assisted synchronization.
\end{abstract}
\maketitle

\section{Introduction} \label{I}
Synchronization is a ubiquitous and compelling collective behavior in nature. Since Huygens first observed synchronization in coupled pendulums in the seventeenth century~\cite{Huygens1673}, systematic research on this phenomenon has expanded across classical and modern physics. Synchronization has subsequently been found in various complex systems, including triode oscillators~\cite{Pikovsky2001}, population dynamics~\cite{Blasius1999}, and cellular metabolism~\cite{PoratShliom2014}, establishing it as a fundamental subject across physics, biology, and engineering~\cite{Aguiar2011,Kuramoto1984,Arenas2008}. This line of research later expanded into the quantum domain. Most early investigations focused on dissipative quantum systems, primarily because conventional synchronization between trajectories relies on the contraction of phase-space volume, as exemplified by qubits coupled to dissipative driven oscillators~\cite{Zhirov2009} and coupled dissipative spins~\cite{Orth2010}. The connection between synchronization and quantum entanglement has also been explored in continuous-variable dissipative systems~\cite{Manzano2013,Mari2013}.

In conservative Hamiltonian systems, however, this standard synchronization mechanism breaks down because Liouville's theorem dictates the conservation of phase-space volume. To address this limitation, Hampton and Zanette introduced the concept of measure synchronization (MS) for coupled Hamiltonian systems~\cite{Hampton1999}. Unlike conventional synchronization, which concerns the locking of instantaneous states or trajectories, MS characterizes synchronization through the statistical measures that the subsystems explore. Specifically, the core principle of MS is to compare the phase-space regions traversed by the trajectories of the two subsystems over long time scales, rather than tracking their instantaneous states~\cite{Ghosh2023,Wang2003,Bemani2017}.

Although exact classical trajectories are strictly absent in quantum mechanics owing to Heisenberg's uncertainty principle, this geometric and statistical perspective of MS can be naturally extended to quantum many-body systems by employing either subsystem expectation-value trajectories and energy expectation-value evolution~\cite{Qiu2014,Sur2020} or quantum phase-space representations such as quasiprobability functions~\cite{PhysRevLett.111.103605}, thereby providing a practical framework for investigating synchronization in conservative quantum regimes.

In recent years, synchronization phenomena in chaotic systems have attracted widespread attention. However, research concerning the connection between quantum chaos and MS in quantum many-body systems remains scarce. Classical chaos is characterized by the exponential sensitivity of trajectories to initial conditions~\cite{Eckmann1985Ergodic}. In contrast, owing to the absence of well-defined phase-space trajectories in quantum many-body systems, chaos cannot be directly characterized by the exponential separation of trajectories as in classical systems~\cite{Casati1979,Eckmann1985}. According to the Bohigas-Giannoni-Schmit (BGS) conjecture, quantum chaos can be identified through spectral statistics~\cite{Bohigas1984a,Bohigas1984b}, providing a theoretical framework for exploring the link between quantum chaos and MS in quantum many-body systems.

On the other hand, entanglement, as a unique nonclassical feature of quantum systems~\cite{Horodecki2009}, has sparked extensive interest in its relationship with quantum chaos~\cite{Neill2016,Piga2019,Rosa2020,Schack1994,Ghose2008,Lee2001,Milonni1983,Furuya1998,Emary2003,Kolovsky2004,Kolovsky2016,Giraldo2020}. In dissipative systems, a connection between synchronization and quantum entanglement has been revealed. In studies of MS in conservative systems, it has also been observed that the onset of MS can be accompanied by pronounced features in the entanglement entropy between subsystems~\cite{Sur2020}. These observations suggest a close connection between MS and quantum entanglement and point to a possible interplay among quantum chaos, MS, and entanglement. However, systematic studies addressing the interplay of these three aspects in quantum many-body systems remain scarce.

To address this gap, this work investigates the interplay between quantum chaos, MS, and quantum entanglement in a two-component ultracold bosonic mixture forming a bosonic Josephson junction. By combining exact diagonalization, which provides the full eigenspectrum and eigenstates of the finite-size Hamiltonian, with spectral-statistics analysis, which serves as a standard diagnostic of quantum chaos, we identify parameter regimes ranging from integrable to deeply chaotic and examine how MS emerges across them. We show that MS can be established at selected quantum chaotic parameter points and that its sensitivity to initial energy imbalance is substantially reduced in the deeply chaotic regime, which may ease initial-state preparation in experiments. In particular, the rapid onset of MS in the deeply chaotic regime is closely tied to the rapid saturation of the bipartite entanglement entropy. Under the eigenstate thermalization hypothesis, chaotic eigenstates behave like random states, so the entanglement entropy of a quantum chaotic system is expected to approach the random-state prediction. As the entanglement entropy quickly approaches this value, the two subsystems also rapidly reach a synchronized state. This rapid MS is established on a much shorter time scale and with stronger entanglement than in the near-chaotic regime. These results provide a systematic account of how quantum chaos, MS, and entanglement are interrelated in a conservative quantum many-body system.

The paper is organized as follows. In Sec.~\ref{II}, we introduce the two-component ultracold bosonic mixture and derive its equivalent spin-representation Hamiltonian. In Sec.~\ref{III}, after eliminating level degeneracies through symmetry analysis, we identify the quantum chaotic parameter regime from spectral statistics and confirm MS from both classical and quantum signatures. In Sec.~\ref{IV}, we compare the initial-state dependence of MS in the near-chaotic and deeply chaotic regimes, and quantitatively examine the correspondence between MS and quantum entanglement to clarify the role of quantum chaotic dynamics in the fast establishment of MS. Finally, Sec.~\ref{V} summarizes the main findings.

\section{Model}\label{II}
To investigate MS in many-body quantum systems, a binary mixture of ultracold bosons confined in a double-well potential serves as an ideal platform known as a two-species bosonic Josephson junction model. Even within the two-mode approximation, the system has an intrinsically complex and nontrivial Hamiltonian. The two bosonic species coexist and tunnel between the two wells while simultaneously undergoing intra-species collisions and inter-species interactions.

Under the two-mode approximation, the many-body Hamiltonian of this coupled system reads~\cite{Milburn1997}
\begin{equation}
\label{e1}
    \hat{H} = \sum_{i,\alpha}
    \left[-\frac{J}{2} \hat{a}^\dagger_{i,\alpha} \hat{a}_{i,\bar{\alpha}} +\frac{U}{2N} \hat{n}_{i,\alpha} (\hat{n}_{i,\alpha} - 1) +\frac{V}{2N} \hat{n}_{i,\alpha} \hat{n}_{\bar{i},\alpha}\right].
\end{equation}
Here, $N$ denotes the particle number of each species, $J$ is the single-particle tunneling amplitude between the two potential wells, and $U$ represents the intra-species interaction strength. The two wells are labeled by $\alpha=L,R$, with $\bar{\alpha}\neq\alpha$ denoting the opposite well, and the two bosonic species are labeled by $i=A,B$, with $\bar{i}\neq i$. The operators $\hat{a}_{i,\alpha}$ and $\hat{a}_{i,\alpha}^{\dagger}$ are the annihilation and creation operators, respectively, for species $i$ in well $\alpha$, and $\hat{n}_{i,\alpha}=\hat{a}_{i,\alpha}^{\dagger}\hat{a}_{i,\alpha}$ is the corresponding particle-number operator. Finally, $V$ characterizes the inter-species interaction, which serves as the sole physical coupling between subsystem $A$ and subsystem $B$ and plays a decisive role in driving the MS between them.

\begin{figure}[hbt!]
	\centering
	\includegraphics[width=1\linewidth]{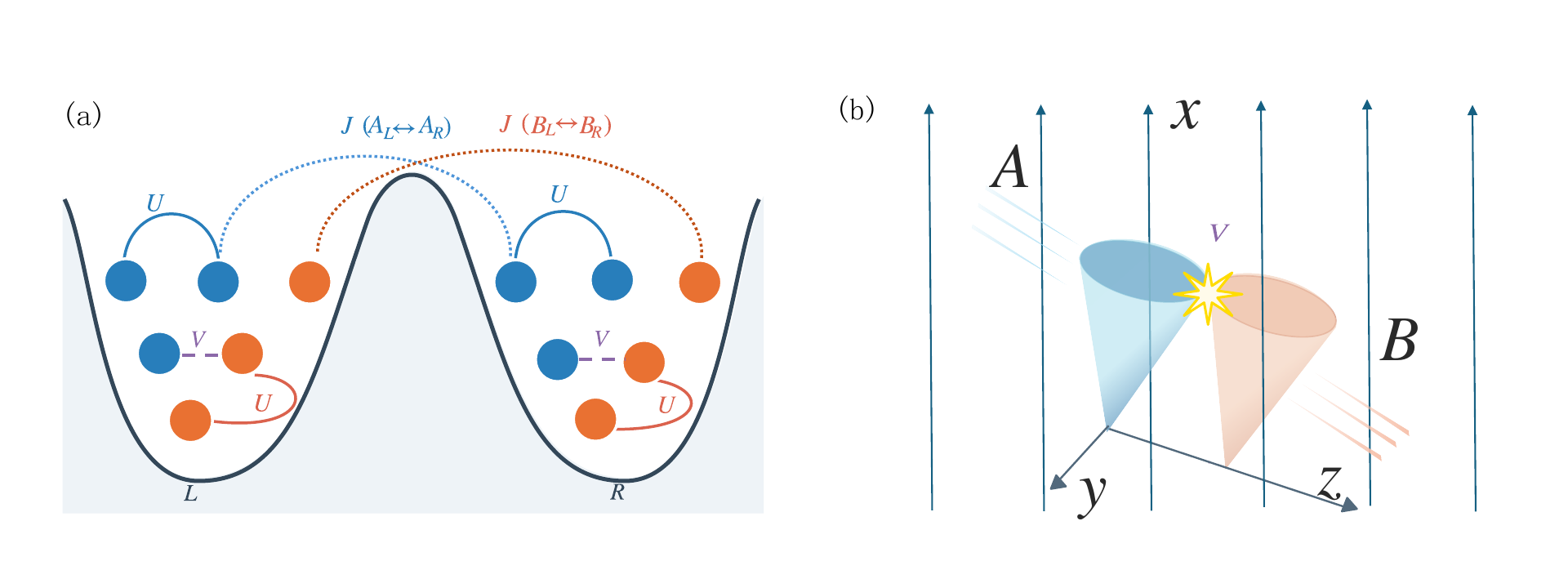}
	\caption{Schematic of the two-component bosonic Josephson junction and its equivalent generalized coupled-top model. (a) Components $A$ (blue) and $B$ (orange) are confined in a double-well potential with wells $L$ and $R$. Each component tunnels between the wells with strength $J$; particles of the same component interact within a well with strength $U$, and the two components interact with strength $V$. (b) The equivalent two macroscopic spins $S$ precessing under the $x$ field and twisting about $z$, coupled via $V\hat{S}_{Az}\hat{S}_{Bz}$.}

   \label{Fig1}
\end{figure}

To gain deeper analytical insight into this many-body complexity, the Hamiltonian in Eq.~(\ref{e1}) can be mapped onto a generalized coupled-top model describing two interacting large spins governed by the angular-momentum algebra. This mapping serves a dual purpose. Mathematically, it aggregates the microscopic bosonic modes into collective macroscopic spin operators via the Schwinger representation, significantly simplifying the Hilbert-space description. Physically, it provides an intuitive geometric picture in the three-dimensional collective spin space that bridges quantum fluctuations and classical nonlinear dynamics, establishing a natural foundation for quantifying synchronization between the two species through the relative orientation of the spin vectors. Each bosonic species thus corresponds to an effective spin of size $S=N/2$, defined through the Schwinger boson representation~\cite{Leggett2001}:
\begin{equation}
\label{e2}
\begin{aligned}
\hat{S}_{ix} &= \frac{1}{2}\left(\hat{a}_{iL}^{\dagger}\hat{a}_{iR}+\hat{a}_{iR}^{\dagger}\hat{a}_{iL}\right),\\
\hat{S}_{iy} &= \frac{i}{2}\left(\hat{a}_{iL}^{\dagger}\hat{a}_{iR}-\hat{a}_{iR}^{\dagger}\hat{a}_{iL}\right),\\
\hat{S}_{iz} &= \frac{1}{2}\left(\hat{n}_{iL}-\hat{n}_{iR}\right),
\end{aligned}
\end{equation}
where these operators satisfy the standard angular-momentum commutation relations $[\hat{S}_{i\alpha},\hat{S}_{j\beta}]=i\delta_{ij}\epsilon_{\alpha\beta\gamma}\hat{S}_{i\gamma}$. Under this representation, the original Hamiltonian in Eq.~(\ref{e1}) can be reduced exactly to~\cite{Robb1998,Emerson2001}
\begin{equation}
\label{e3}
\hat{H}=\hat{H}_{A}+\hat{H}_{B}+\hat{H}_{AB},
\end{equation}
with the individual and coupling terms given by
\begin{equation}
\label{e4}
\begin{aligned}
\hat{H}_{A} &= -J\hat{S}_{Ax}+\frac{U_A}{2S}\hat{S}_{Az}^{2},\\
\hat{H}_{B} &= -J\hat{S}_{Bx}+\frac{U_B}{2S}\hat{S}_{Bz}^{2},\\
\hat{H}_{AB} &= \frac{V}{S}\hat{S}_{Az}\hat{S}_{Bz},
\end{aligned}
\end{equation}
where $U_A$ and $U_B$ denote the intra-species interaction strengths for species $A$ and $B$, respectively. A detailed derivation is provided in Appendix~\ref{AA}. Energy and time are scaled in units of $J$ and $1/J$, respectively.

In what follows, the exact quantum dynamics of the system is governed by the Schrödinger equation with Hamiltonian~(\ref{e3}) in the full Hilbert space of the two collective spins. This treatment fully accounts for the evolution of pseudospin expectation values, relative quantum fluctuations, and quantum entanglement between the two bosonic species, all of which are essential for rigorously evaluating quantum MS.

\section{Identifying Measure Synchronization}\label{III}

\subsection{Spectral Statistics}\label{IIIA}

\begin{figure}[hbt!]
	\centering
	\includegraphics[width=1\linewidth]{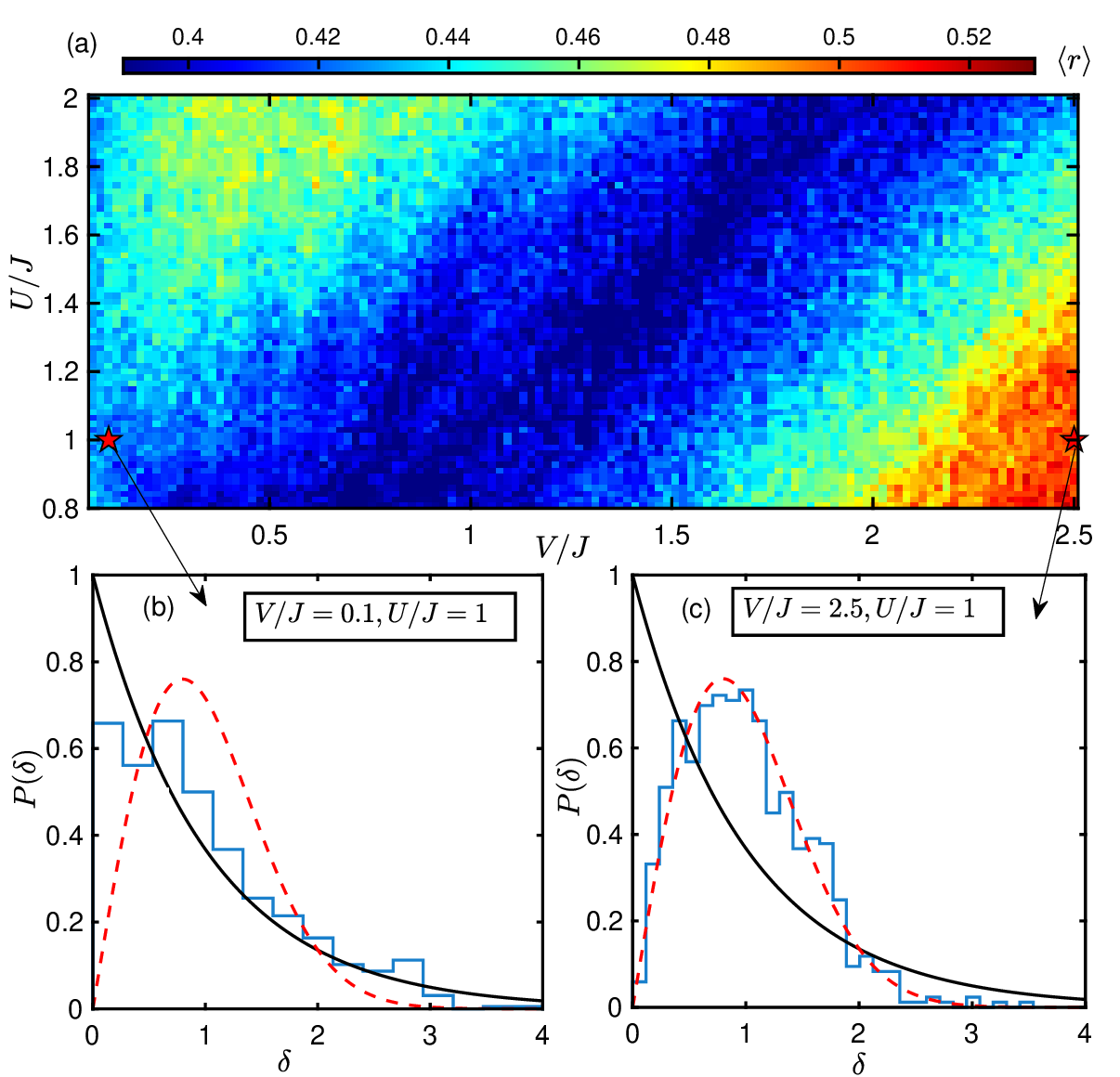}
	\caption{(a) Colormap of the average level-spacing ratio $\langle r\rangle$ in the $U$--$V$ parameter plane, showing the transition between the integrable (blue) and quantum chaotic (red) regimes. Parameters: $U_A=U_B=U$ and $N_A=N_B=N=2S=40$. Nearest-neighbor level-spacing distributions $P(\delta)$ (histograms) for the marked (b) integrable and (c) chaotic points in (a), where $\delta$ is in units of $J$. The Poisson distribution $P_{\mathrm{P}}(\delta)$ (black solid line) and the Wigner-Dyson distribution $P_{\mathrm{WD}}(\delta)$ (red dashed line) are included for comparison. }
	\label{Fig2}
\end{figure}

To explore MS in the quantum chaotic regime, we first identify the parameter regimes in which chaos emerges, thereby providing a robust basis for investigating synchronization dynamics. Signatures of quantum chaos are commonly diagnosed through the spectral statistics of the underlying Hamiltonian. To this end, we examine the eigenenergies $\mathcal{E}_n$ and eigenstates
\begin{equation}
\vert\psi_n\rangle=\sum_{i=1}^{\mathcal{N}}\psi_i^{(n)}\vert m_A^{(i)},m_B^{(i)}\rangle,
\end{equation}
obtained by exact diagonalization of the Hamiltonian in Eq.~(\ref{e3}), where $m_A^{(i)}$ and $m_B^{(i)}$ denote the eigenvalues of $\hat{S}_{Az}$ and $\hat{S}_{Bz}$, respectively, and $\mathcal{N}=(N_A+1)(N_B+1)$ denotes the dimension of the Hilbert space.

When the parameters in both subsystems are identical, the Hamiltonian in Eq.~(\ref{e3}) possesses two exact discrete symmetries that constrain the energy spectrum structure. The first is the spatial inversion (parity) symmetry corresponding to the interchange of the left and right potential wells, which leads to the sign reversal of the operators $\hat{S}_{iy}$ and $\hat{S}_{iz}$ and is described by the operator $\hat{\Pi} = e^{i\pi(\hat{S}_{Ax}+\hat{S}_{Bx})}$. The second is the species-exchange symmetry represented by the operator $\hat{O}$ that swaps the two subsystems such that $\hat{O}\vert{}m_A, m_B\rangle = \vert{}m_B, m_A\rangle$. Since both operators commute with the Hamiltonian and with each other,
\begin{equation}
\label{B1}
[\hat{\Pi}, \hat{H}] = [\hat{O}, \hat{H}] = [\hat{\Pi}, \hat{O}] = 0,
\end{equation}
they share simultaneous eigenvalues of $\pm 1$. Consequently, the Hilbert space and eigenmodes can be rigorously classified into four distinct symmetry sectors:
\begin{itemize}
\item even-even sector: $\langle\psi_n\vert{}\hat{\Pi}\vert{}\psi_n\rangle = 1$, $\langle\psi_n\vert{}\hat{O}\vert{}\psi_n\rangle = 1$;
\item even-odd sector: $\langle\psi_n\vert{}\hat{\Pi}\vert{}\psi_n\rangle = 1$, $\langle\psi_n\vert{}\hat{O}\vert{}\psi_n\rangle = -1$;
\item odd-even sector: $\langle\psi_n\vert{}\hat{\Pi}\vert{}\psi_n\rangle = -1$, $\langle\psi_n\vert{}\hat{O}\vert{}\psi_n\rangle = 1$;
\item odd-odd sector: $\langle\psi_n\vert{}\hat{\Pi}\vert{}\psi_n\rangle = -1$, $\langle\psi_n\vert{}\hat{O}\vert{}\psi_n\rangle = -1$.
\end{itemize}

We restrict our analysis to the even-even sector to avoid the artificial mixing of independent spectra and accidental level crossings, which would otherwise distort the spectral statistics; other sectors exhibit the same universal statistical behavior. The sorted eigenenergies in this sector are used to construct the adjacent level-spacing ratio~\cite{Mondal2020Chaos,Atas2013,Oganesyan2007}
\begin{equation}
\label{e5}
r_n = \min\left\{ \frac{\delta_n}{\delta_{n+1}}, \frac{\delta_{n+1}}{\delta_n} \right\},
\end{equation}
where $\delta_n = \mathcal{E}_n - \mathcal{E}_{n-1}$. 
By averaging $r_n$ over the energy spectrum within this sector, we obtain the mean level-spacing ratio $\langle r \rangle=\frac{1}{\mathcal{N}-2}\sum_{n=2}^{\mathcal{N}-1} r_n$.
As a standard diagnostic, this ratio approaches the Gaussian Orthogonal Ensemble (GOE) limit with $\langle r \rangle_{\text{GOE}} \approx 0.530$ in the fully chaotic regime due to level repulsion, whereas it follows the Poisson limit with $\langle r \rangle_{\text{PS}} \approx 0.386$ for generic integrable systems~\cite{Atas2013,Haake2010}. 
Figure~\ref{Fig2}(a) shows the distribution of $\langle r \rangle$ across the $U$--$V$ parameter plane, revealing a clear progression of dynamical regimes. Blue regions, where the inter-species interaction $V$ is relatively weak compared to the intra-species interaction $U$, correspond to integrable behavior governed by Poisson statistics, whereas red regions, where $V$ strongly dominates, indicate chaotic behavior governed by GOE statistics, with an intermediate regime bridging these distinct limits.

These chaotic signatures are further corroborated by analyzing the nearest-neighbor level-spacing distribution $P(\delta)$. According to the Berry-Tabor conjecture, integrable systems exhibit Poissonian level statistics, whereas the Bohigas-Giannoni-Schmit (BGS) conjecture establishes that quantum chaotic systems follow the Wigner-Dyson distribution characterized by prominent level repulsion~\cite{Berry1977,Bohigas1984b}. Evaluating $P(\delta)$ at representative parameter points confirms this dichotomy. In the chaotic regime [Fig.~\ref{Fig2}(b)], the distribution exhibits excellent agreement with the Wigner-Dyson prediction $P_{\mathrm{WD}}(\delta)=\frac{\pi\delta}{2}\exp\left(-\frac{\pi\delta^2}{4}\right)$, whereas in the integrable regime [Fig.~\ref{Fig2}(c)], the distribution strictly adheres to Poisson statistics $P_{\mathrm{P}}(\delta)=\exp(-\delta)$.

\subsection{Measure Synchronization}\label{IIIB}

\begin{figure}[hbt!]
	\centering
      \begin{adjustbox}{trim=3cm 2cm 3cm 2cm, clip}
	\includegraphics[width=1.5\linewidth]{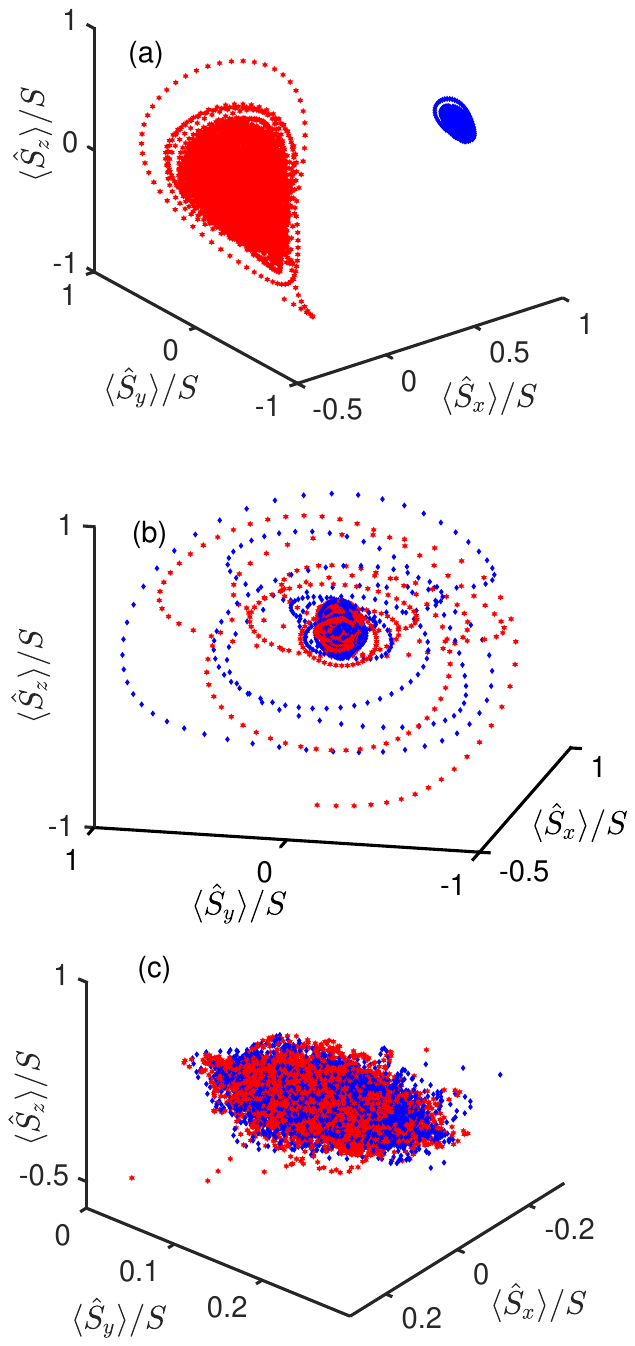}
    \end{adjustbox}
    \caption{Phase-space characterization of MS in the quantum system. The trajectory traced by each subsystem during the evolution is tracked by the normalized spin expectation vector $(\langle\hat{S}_x\rangle,\langle\hat{S}_y\rangle,\langle\hat{S}_z\rangle)/S$ in the three-dimensional collective spin space. The trajectories of species $A$ and $B$ are represented by blue diamonds and red hexagons, respectively, with each data point corresponding to a discrete sampling instant during the evolution. (a) $V/J=0.1$ (non-MS). The phase-space regions covered by the two subsystems are separated with no overlap. (b) and (c) correspond to MS in the near-chaos regime ($V/J=1.6$) and chaotic regimes ($V/J=2.5$), respectively. The phase-space regions covered by the two subsystems completely coincide. The other parameters are $S=20$ and $U/J=1$.}
	\label{Fig4}
\end{figure}

Building on the static spectral characterization based on the level-spacing ratio $\langle r\rangle$ and the level-spacing distribution $P(\delta)$, we now investigate the MS dynamics across distinct parameter regimes. In classical nonlinear dynamics, the fundamental criterion for MS is that the invariant measures or phase-space regions explored by the two interacting subsystems coincide completely during long-term evolution~\cite{Wang2003b}. In quantum mechanics, however, quantum uncertainty prevents a system from possessing precise phase-space point trajectories in the classical sense. To rigorously characterize measure synchronization in the quantum many-body system examined here, we analyze the synchronization behavior from three complementary perspectives. (i) We examine the trajectories traced by the normalized spin expectation vectors of the two subsystems on the three-dimensional Bloch-like space. (ii) We track the temporal evolution of the subsystem energy expectation values to monitor the sharing and convergence of energy between the two subsystems. (iii) We analyze the temporal evolution of spin quantum fluctuations within each subsystem, which allows us to evaluate whether the statistical distributions and underlying quantum measures become fully aligned.

We implement the first criterion by defining the normalized spin expectation vectors for subsystems $A$ and $B$ as $(\langle\hat{S}_{Ax}\rangle/S_A,\langle\hat{S}_{Ay}\rangle/S_A,\langle\hat{S}_{Az}\rangle/S_A)$ and $(\langle\hat{S}_{Bx}\rangle/S_B,\langle\hat{S}_{By}\rangle/S_B,\langle\hat{S}_{Bz}\rangle/S_B)$, respectively, and examining their trajectories in the three-dimensional collective spin space. The initial state $\vert\psi(0)\rangle$ is chosen as the product state of spin coherent states for the two subsystems $A$ and $B$, such that $\vert\psi(0)\rangle=\vert z_1,\phi_1\rangle\otimes\vert z_2,\phi_2\rangle$. The single-component spin coherent state $\vert z,\phi\rangle$ is defined as~\cite{Radcliffe1971}
\begin{equation}
\label{e7}
|z,\phi\rangle=\left(\frac{1+z}{2}\right)^S\exp\left(\frac{1-z}{1+z}e^{i\phi}\hat{S}^-\right)|S,S\rangle.
\end{equation}
Unless otherwise stated, we set the initial parameters to $z_1=0,\phi_1=0$ and $z_2=1,\phi_2=0$ in all following simulations. Crucially, these choices place the two subsystems in strikingly disparate initial configurations.Subsystems $A$ and $B$ are initialized as spin coherent states pointing along the positive $x$-axis on the equator and the north pole, respectively. This configuration places subsystem $A$ in a regime where tunneling energy significantly dominates, whereas subsystem $B$ resides in a regime dominated by self-interaction energy. Initiating the temporal evolution from such markedly contrasting initial states provides a stringent and rigorous test of whether chaotic dynamics has a robust capacity for convergence and synchronization, thereby making the emergence of measure synchronization particularly compelling.

By tracking the evolution of this state across the integrable, intermediate, and quantum chaotic regimes, we examine how MS dynamics emerges and evolves at these representative parameter points. As shown in Fig.~\ref{Fig4}(a), the regions of the three-dimensional collective spin phase space explored by the two subsystems during their evolution remain completely separated in the integrable regime ($V/J=0.1$ and $U/J=1$), reflecting typical non-MS behavior. In the intermediate inter-species interaction regime ($V/J=1.6$ and $U/J=1$) [Fig.~\ref{Fig4}(b)], the phase-space regions traversed by the two subsystems begin to overlap significantly, yet regular and separated trajectory structures remain clearly discernible alongside a localized densely interwoven domain, representing a transitional form of MS in this near-chaos parameter regime. By contrast, at the fully chaotic parameter point ($V/J=2.5$ and $U/J=1$) [Fig.~\ref{Fig4}(c)], the phase-space regions almost coincide completely, with their trajectories densely interwoven within the same geometric domain and becoming statistically indistinguishable. This feature provides a direct signature of MS, showing that MS can emerge reliably in the quantum chaotic regime despite disparate initial configurations. It should be noted, however, that this observation does not imply that chaos alone guarantees MS, as the chaotic parameters must still be selected appropriately for complete synchronization to manifest.

\begin{figure}[hbt!]
\centering
\includegraphics[width=1\linewidth]{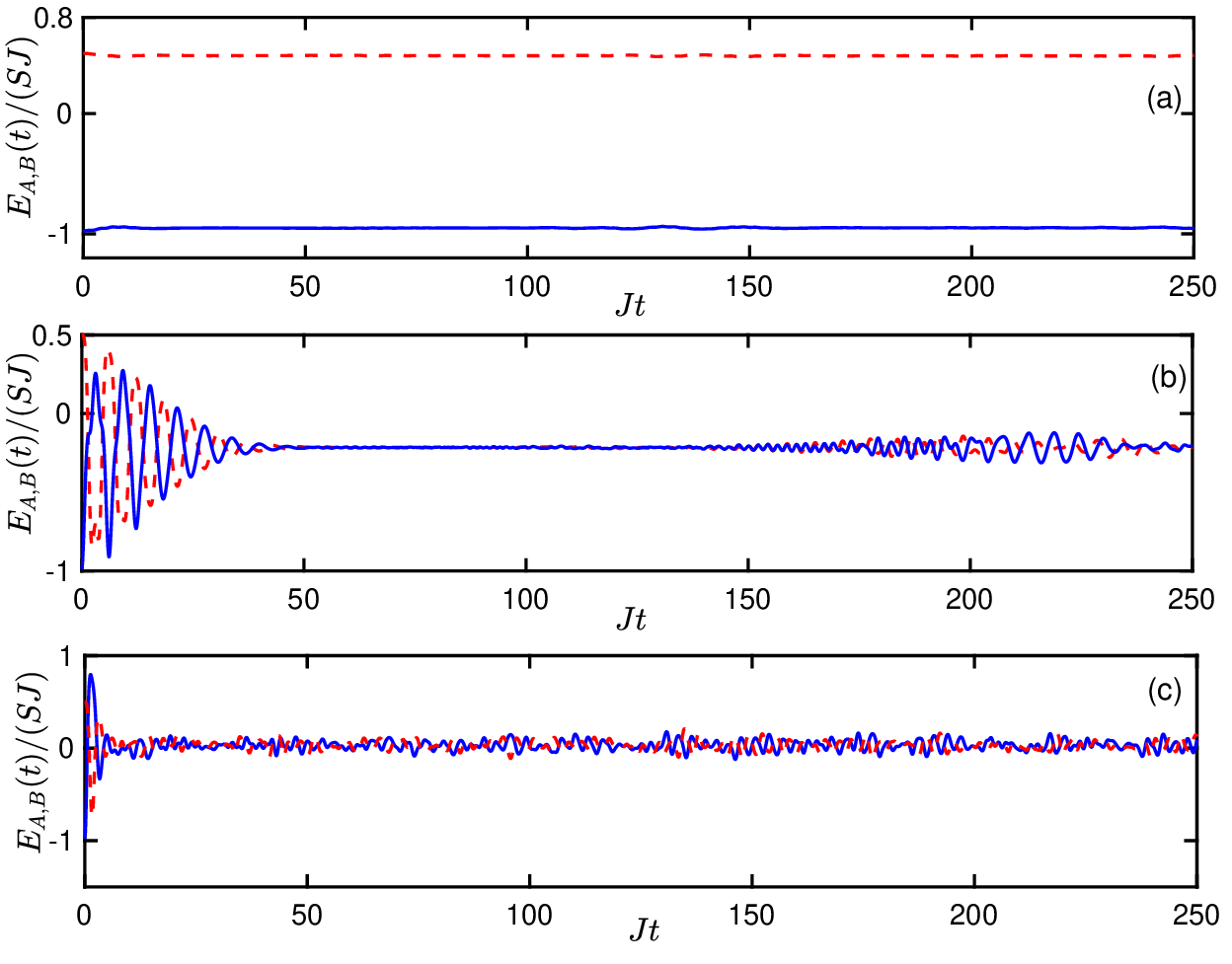}
\caption{Time evolution of the normalized energy expectation values $E_A(t)/(SJ)$ and $E_B(t)/(SJ)$ for subsystems $A$ and $B$. Panels (a), (b), and (c) correspond to the integrable regime with $V/J=0.1$, the near-chaos regime with $V/J=1.6$, and the deeply chaotic regime with $V/J=2.5$, respectively. Subsystems $A$ and $B$ are represented by blue solid lines and red dashed lines, respectively. Parameters are $S=20$ and $U/J=1$.}
\label{Fig5}
\end{figure}

We further examine the real-time evolution of the subsystem energy expectation values, $E_A(t)=\langle\psi(t)\vert{}\hat{H}_A\vert{}\psi(t)\rangle$ and $E_B(t)=\langle\psi(t)\vert{}\hat{H}_B\vert{}\psi(t)\rangle$, to provide energetic evidence for MS. Figure~\ref{Fig5} presents the energy dynamics across the integrable, intermediate, and fully chaotic regimes. In the non-MS integrable case [$V/J=0.1$, Fig.~\ref{Fig5}(a)], the energy trajectories of the two subsystems remain distinctly separated throughout the entire temporal window, reflecting constrained energy redistribution and the persistence of the initial energy asymmetry. The negative value of $E_A(t)$ reflects the dominance of the tunneling energy in the initial state of subsystem A. In the intermediate near-chaos regime [$V/J=1.6$, Fig.~\ref{Fig5}(b)], the two subsystems undergo a prolonged transient stage of fast oscillations before their energy expectation values gradually converge and fluctuate around a common long-time average. By contrast, at the representative fully chaotic MS point [$V/J=2.5$, Fig.~\ref{Fig5}(c)], the two subsystems undergo rapid and intense energy exchange starting immediately from the early transient stage. Despite beginning with drastically different initial energy configurations, their respective energy expectation values quickly converge and subsequently oscillate synchronously around a common average. This dynamical energy equilibration and sharing indicate that inter-subsystem coupling enables effective energy exchange and eliminates initial energetic disparities, providing a clear energetic counterpart to the phase-space MS across different parameter regimes.

\begin{figure}[hbt!]
\centering
\includegraphics[width=1.05\linewidth]{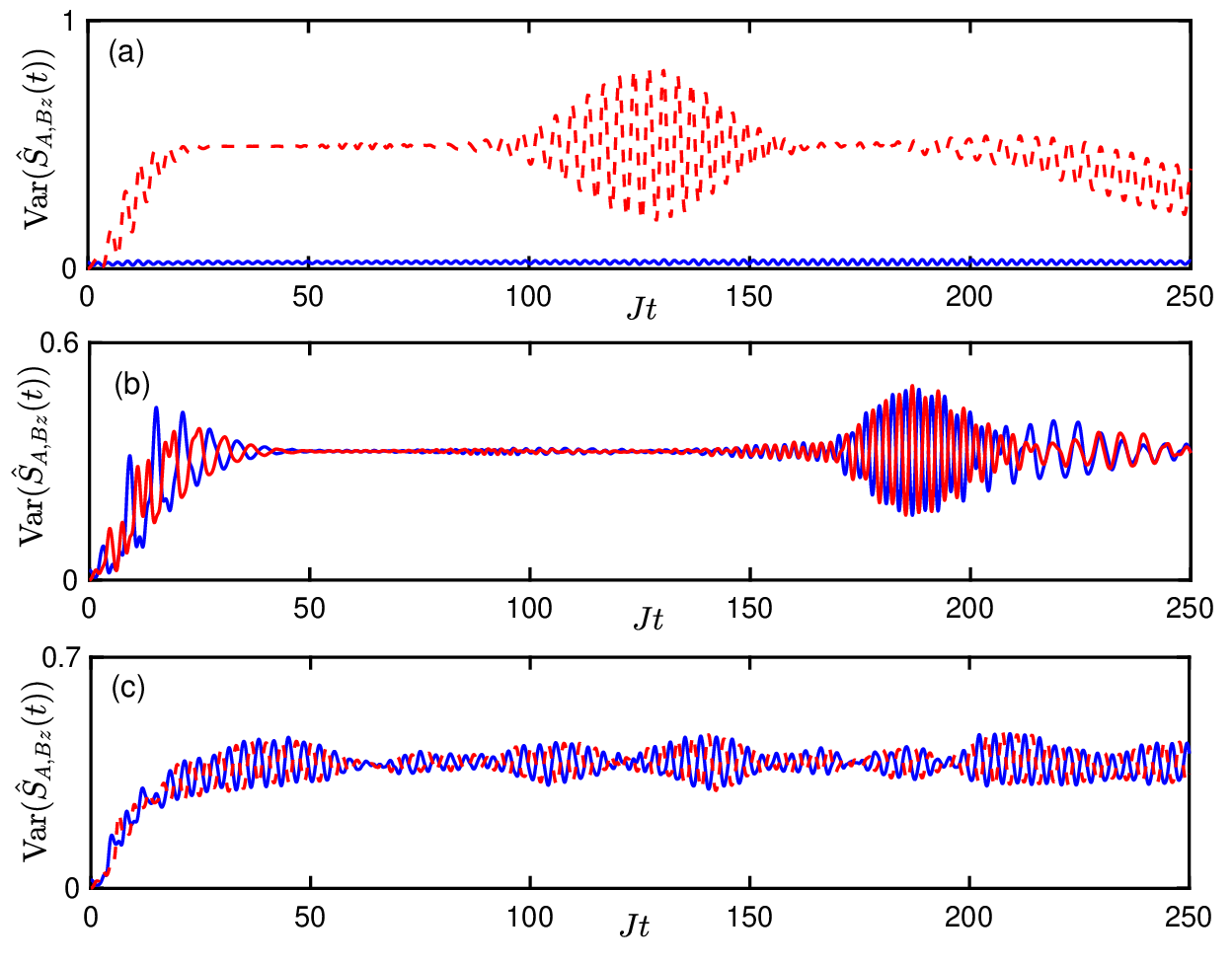}
\caption{Temporal evolution of the normalized spin fluctuation $\mathrm{Var}(\hat{S}_{iz}(t))/S^2$ for subsystems $A$ (blue solid lines) and $B$ (red dashed lines) across three dynamic regimes. Panels (a), (b), and (c) correspond to the integrable regime with $V/J=0.1$, the near-chaos regime with $V/J=1.6$, and the deeply chaotic regime with $V/J=2.5$, respectively. The parameters are the same as in the previous figures.}
\label{FigMB}
\end{figure}

We then investigate the quantum fluctuations of $\hat{S}_{iz}$, which quantify the double-well population-imbalance fluctuations within each subsystem, to explore the dynamical evolution of their quantum statistical properties. Figure~\ref{FigMB} shows the temporal evolution of the normalized variance of these fluctuations, $\mathrm{Var}(\hat{S}_{iz}(t))=[\langle\hat{S}_{iz}^2\rangle-\langle\hat{S}_{iz}\rangle^2]/S^2$, for the two subsystems $i=A,B$ across the integrable, near-chaotic, and deeply chaotic regimes. In the integrable regime [$V/J=0.1$, Fig.~\ref{FigMB}(a)], the fluctuations of subsystems $A$ (blue solid lines) and $B$ (red dashed lines) both exhibit high-frequency oscillations modulated by slowly varying envelopes. Although the two subsystems are governed by dynamical equations of the same form, their envelope amplitudes, detailed time series, and long-time averages differ markedly, indicating a lack of cooperative statistical correlation. In the near-chaotic regime [$V/J=1.6$, Fig.~\ref{FigMB}(b)], the fluctuations undergo a prolonged transient stage before gradually approaching each other, eventually settling into persistent small-amplitude oscillations around a nearly identical mean value. By contrast, in the deeply chaotic regime [$V/J=2.5$, Fig.~\ref{FigMB}(c)], after a brief transient stage, the fluctuation dynamics of the two subsystems become remarkably anti-phase-locked with symmetric oscillation amplitudes, and the discrepancy between their long-time average values vanishes completely. This robust anti-phase correlation reflects a complementary sharing of quantum fluctuations between the two coupled subsystems. The evolution across these regimes shows that, with increasing chaoticity, the statistical alignment between the two subsystems changes from slow convergence to rapid and complete anti-phase locking, confirming that measure synchronization extends beyond classical expectation values down to the level of quantum fluctuations.

The analysis across three complementary perspectives provides a rigorous validation of MS in the quantum system. Although chaotic dynamics does not universally guarantee MS, our findings demonstrate that within representative chaotic parameter regimes, chaos efficiently bridges disparate initial configurations. The coincidence of phase-space trajectories, the equilibration of subsystem energies, and the synchronization of population imbalance fluctuations collectively confirm that MS achieves full statistical alignment. These results establish MS as a robust collective phenomenon, offering deeper insights into quantum dynamics of interacting many-body systems in Chaos.

\section{Measure Synchronization in Quantum Chaotic Regimes}\label{IV}

\subsection{Initial-State Independence}\label{IVC}

\begin{figure}[hbt!]
\centering
\includegraphics[width=1\linewidth]{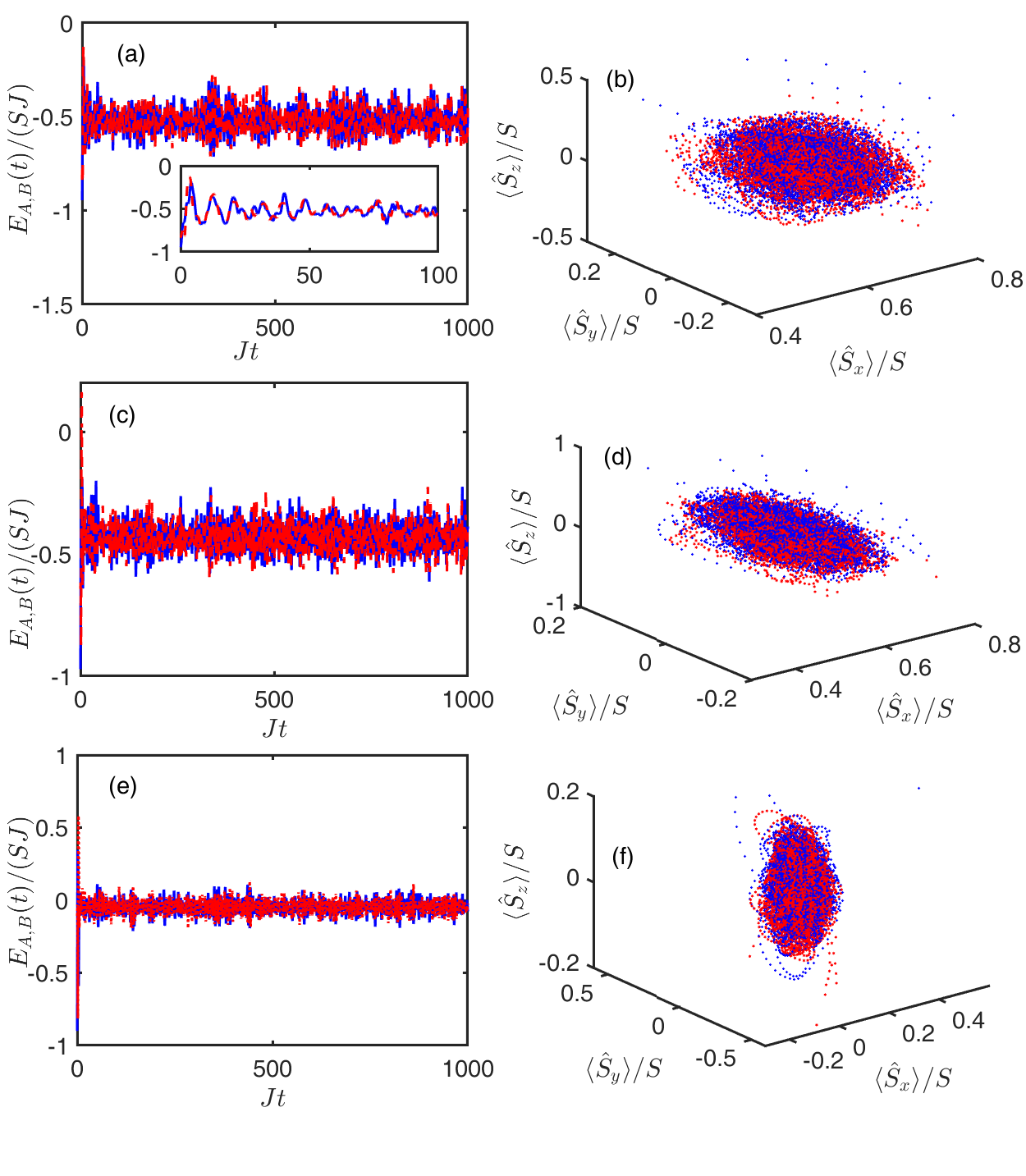}
\caption{Measure synchronization dynamics for three different initial energy states in deeply chaotic regime. Panels (a,b), (c,d), and (e,f) correspond to initial states with distinct energy imbalances: $(z_1,\phi_1;z_2,\phi_2)=(0.2,0;0.4,0)$, $(0.1,0;0.6,0)$, and $(0.3,0;0.9,0)$, respectively. Left column (a,c,e): time evolution of the energy expectation values $E_A(t)$ and $E_B(t)$ of the two subsystems, with panel (a) including a zoom-in of the early-time transient stage. Right column (b,d,f) depicts the phase-space distributions formed by the expectation values of species $A$ (blue diamonds) and $B$ (red hexagons) sampled at fixed time intervals during the evolution. Parameters are $S=20$, $U/J=0.8$, and $V/J=2.4$.}
\label{Fig_Appendix.eps}
\end{figure}

While the preceding sections confirm the emergence of MS in the chaotic regime, a fundamental question remains regarding whether this collective phenomenon relies on specific, carefully prepared initial states. Ordinarily, chaotic dynamics is characterized by an extreme sensitivity to initial conditions, where diverging trajectories typically prevent any convergence between distinct initial configurations. However, MS exhibits a striking robustness against variations in initial states. To examine this resilience, we select three distinct product states of spin coherent states with significantly different initial energy imbalances, namely $(z_1, \phi_1; z_2, \phi_2) = (0.2, 0; 0.4, 0)$, $(0.1, 0; 0.6, 0)$, and $(0.3, 0; 0.9, 0)$. All numerical simulations are performed within the deeply chaotic parameter regime. As shown in Fig.~\ref{Fig_Appendix.eps}, initial states with varying energy imbalances evolving in the deeply chaotic region consistently exhibit dynamical characteristics identical to MS. Specifically, the $(S_x, S_y, S_z)$ phase-space regions explored by the expectation values of each subsystem overlap completely, and the two subsystems undergo mutual energy exchange, eventually oscillating around a common long-time average.

\begin{figure}[hbt!]
\centering
\includegraphics[width=1\linewidth]{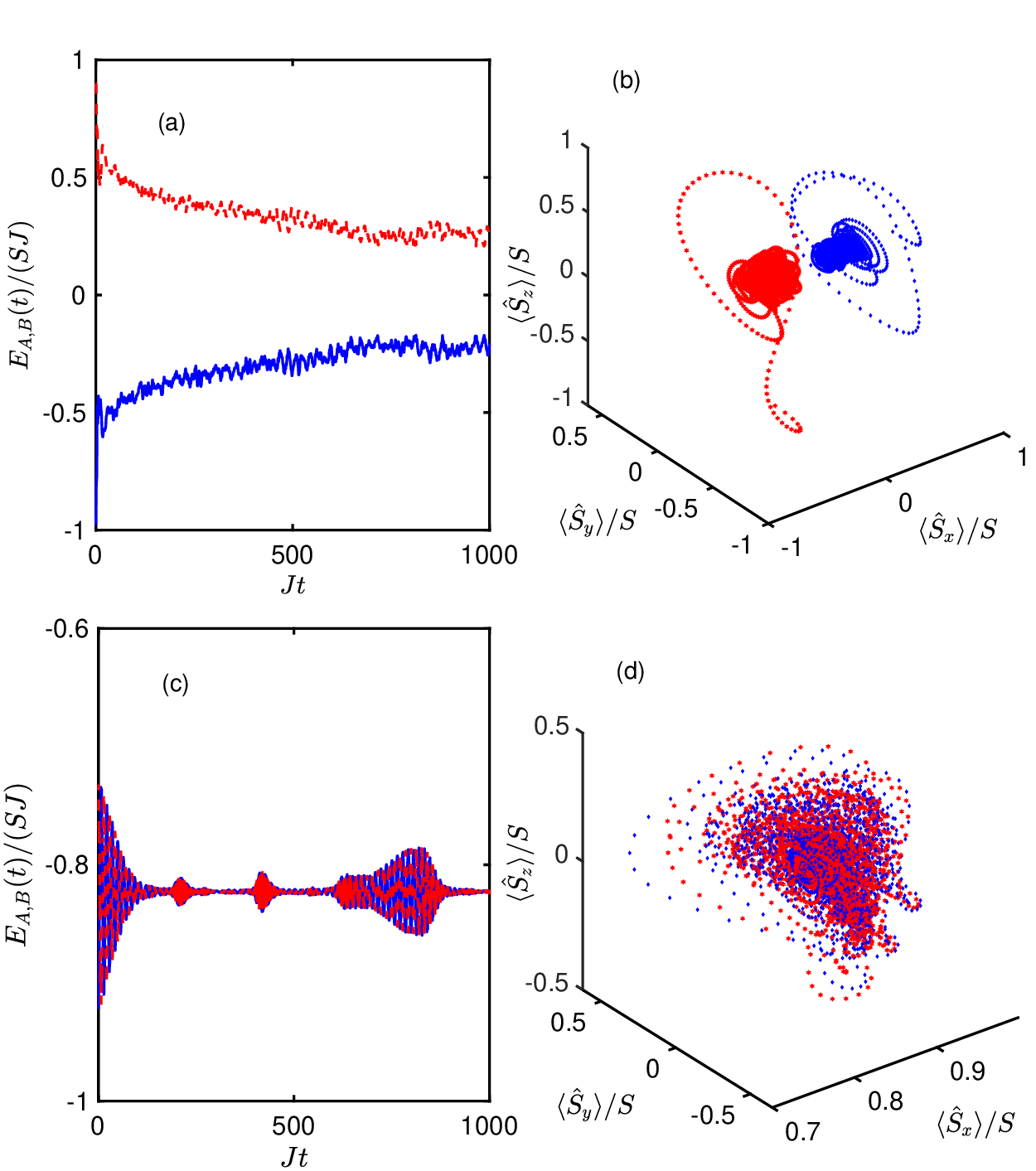}
\caption{Initial-state dependence of the MS dynamics in the near-chaotic regime. Panels (a,b) and (c,d) correspond to initial states with distinct energy imbalances: $(z_1,\phi_1;z_2,\phi_2)=(0,0;1,0)$ and $(0.2,0;0.4,0)$, respectively. Left column (a,c): time evolution of the energy expectation values $E_A(t)$ and $E_B(t)$ of the two subsystems. Right column (b,d) depicts the phase-space distributions formed by the expectation values of species $A$ (blue diamonds) and $B$ (red hexagons) sampled at fixed time intervals during the evolution. Parameters are $S=20$, $U/J=1.8$, and $V/J=0.7$.}
\label{Fig6.eps}
\end{figure}

For comparison, we evaluate the initial-state dependence of MS in the near-chaos regime in Fig.~\ref{Fig6.eps}, where for identical Hamiltonian parameters $V/J=0.7$ and $U/J=1.8$, two distinct initial states with different energy imbalances, namely $(z_1,\phi_1;z_2,\phi_2)=(0,0;1,0)$ and $(0.2,0;0.4,0)$, exhibit fundamentally different dynamical behaviors. For the first initial state, MS fails to occur. Although the dynamical trajectories of subsystems $A$ and $B$ enter chaotic regimes, they remain confined to two isolated phase-space regions, with their corresponding energy expectation values $E_A(t)$ and $E_B(t)$ remaining distinctly separated throughout the temporal evolution. For the second initial state, the rapid convergence of energy expectation values together with the strongly overlapping phase-space trajectories clearly indicates the emergence of MS. This contrast reveals that unlike the robust, initial-state-independent MS observed in the deeply chaotic regime, MS in the near-chaos regime exhibits strong sensitivity to initial conditions due to the coexisting regular and chaotic phase-space structures.

As the inter-species interaction strength increases into the deeply chaotic regime, this sensitive initial-state dependence gives way to a highly robust initial-state independence. Although chaotic dynamics typically amplifies sensitivity to initial conditions and generates divergent individual evolution paths, the expectation-value trajectories of subsystems $A$ and $B$ within the deeply chaotic regime do not permanently separate. Instead, they intertwine and wrap around each other within the same invariant region of phase space. Specifically, the powerful global phase-space mixing mechanism guarantees that, regardless of disparate initial configurations, the expectation values of both subsystems are driven to explore identical phase-space volumes, establishing universal and robust MS.

From an experimental perspective, this initial-state robustness significantly alleviates the stringent demands for precise initial state preparation in ultracold atomic platforms, confirming that MS can be reliably achieved from arbitrary initial energy configurations. Consequently, this dynamical resilience establishes MS as a robust collective phenomenon, providing a reliable pathway for sharing dynamical properties between interacting quantum components.

\subsection{Entanglement and Measure Synchronization}

Building on the preceding section, in which the initial energy imbalance and the evolution of expectation values showed that MS at the deeply chaotic regime is remarkably independent of the initial state, we now address a deeper question regarding the quantum-information mechanisms underlying this robust sharing behavior. Specifically, we ask whether this collective dynamical behavior is accompanied by specific quantum-correlation structures and whether an intrinsic link exists among MS, quantum entanglement, and quantum chaos. To address this question, we quantitatively evaluate entanglement dynamics by computing the time evolution of the bipartite entanglement entropy $S_{\mathrm{en}}(t)$ between subsystems $A$ and $B$, defined as
\begin{equation}
\label{e6}
S_{\mathrm{en}}(t)=-\operatorname{Tr}[\hat{\rho}_B(t)\ln\hat{\rho}_B(t)],
\end{equation}
where $\hat{\rho}_B(t)=\operatorname{Tr}_A|\psi(t)\rangle\langle\psi(t)|$ is the reduced density matrix of subsystem $B$ obtained by tracing over subsystem $A$. Here $|\psi(t)\rangle=e^{-i\hat{H}t}|\psi(0)\rangle$ is the time-evolved state, with the initial state $|\psi(0)\rangle$ prepared as a tensor product of spin coherent states.

\begin{figure}[hbt!]
\centering
\includegraphics[width=1\linewidth]{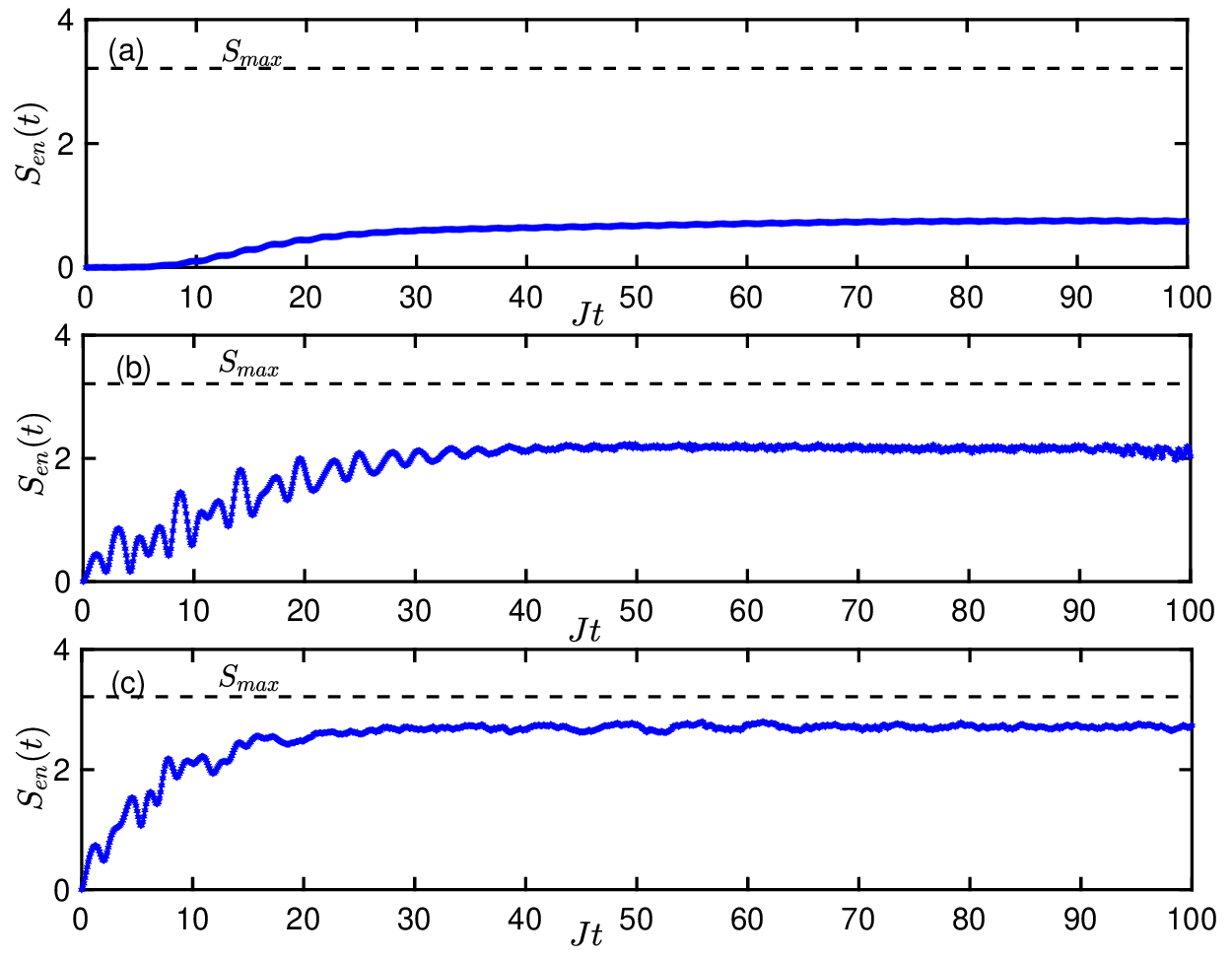}
\caption{Time evolution of the bipartite entanglement entropy $S_{\mathrm{en}}(t)$ between subsystems $A$ and $B$. (a) Integrable regime with $V/J=0.05$. (b) Near-chaotic regime with $V/J=1.6$. (c) Deeply chaotic regime with $V/J=2.5$. The black dashed line in each panel marks the random-state prediction $S_{\max}$. The initial state is a direct product of spin coherent states of the two subsystems. The initial state is chosen as $(z_1, \phi_1; z_2, \phi_2) = (0.01, 0; 0.8, 0)$. The parameters are $S=20$ and $U/J=1$.}
\label{Fig3}
\end{figure}

The temporal evolution of $S_{\mathrm{en}}(t)$ exhibits distinct characteristics across the integrable, near-chaotic, and deeply chaotic regimes. In the integrable regime [$V/J=0.05$, Fig.~\ref{Fig3}(a)], robust local constants of motion and regular phase-space structures inhibit global ergodicity, which strongly suppresses entanglement growth and keeps the entropy far below the theoretical maximum $S_{\max}=\ln(2S+1)-1/2$ predicted for random states in finite-dimensional Hilbert spaces~\cite{Page1993}, where each subsystem has Hilbert-space dimension $2S+1$. In the near-chaotic regime [$V/J=1.6$, Fig.~\ref{Fig3}(b)], the entanglement entropy undergoes a prolonged transient stage with slow oscillations and eventually saturates at a value noticeably higher than that in the integrable regime, yet still well below $S_{\max}$. By contrast, in the deeply chaotic regime [$V/J=2.5$, Fig.~\ref{Fig3}(c)], the entanglement entropy grows rapidly from zero during the early transient stage and quickly reaches near-maximal saturation close to $S_{\max}$. This rapid saturation is consistent with quantum chaos and the eigenstate thermalization hypothesis, according to which chaotic many-body dynamics drives the system toward a statistically random state within the accessible Hilbert space.

Building on these dynamical features, we now investigate how the long-time stabilization of quantum entanglement correlates with the emergence of MS for varying inter-species interaction strengths $V/J$. A direct and observable consequence of MS behavior is that the long-time average energies of the two subsystems tend to become equal. To quantitatively characterize the degree of measure synchronization, we introduce the normalized difference between the long-time averaged energies of the two subsystems together with the long-time averaged energy of each subsystem,
\begin{equation}
\label{eq9}
\begin{aligned}
\langle\Delta\rangle &= \frac{|\langle E_A\rangle-\langle E_B\rangle|}{SJ},\\
\langle E_{A(B)}\rangle &= \lim_{T\to\infty}\frac{1}{T}\int_0^T E_{A(B)}(t)\,dt,
\end{aligned}
\end{equation}
where $\langle E_A\rangle$ and $\langle E_B\rangle$ denote the long-time averaged energies of subsystems $A$ and $B$, respectively. This quantity directly measures the convergence of the subsystem average energies, with $\langle\Delta\rangle\to0$ signifying the realization of MS. To further evaluate the parameter dependence of quantum correlations, we define the ratio of the long-time average entanglement entropy to the maximum random-state entanglement entropy, $\langle S_{\mathrm{en}}\rangle/S_{\max}$, where the long-time average is calculated as
\begin{equation}
\label{eq10}
\langle S_{\mathrm{en}}\rangle=\lim_{T\to\infty}\frac{1}{T}\int_0^T S_{\mathrm{en}}(t)\,dt.
\end{equation}
This ratio characterizes the steady-state strength of quantum correlations. A value approaching unity indicates maximal entanglement between the subsystems, whereas a value significantly below unity reflects weak correlations.

\begin{figure}[hbt!]
\includegraphics[width=1\linewidth]{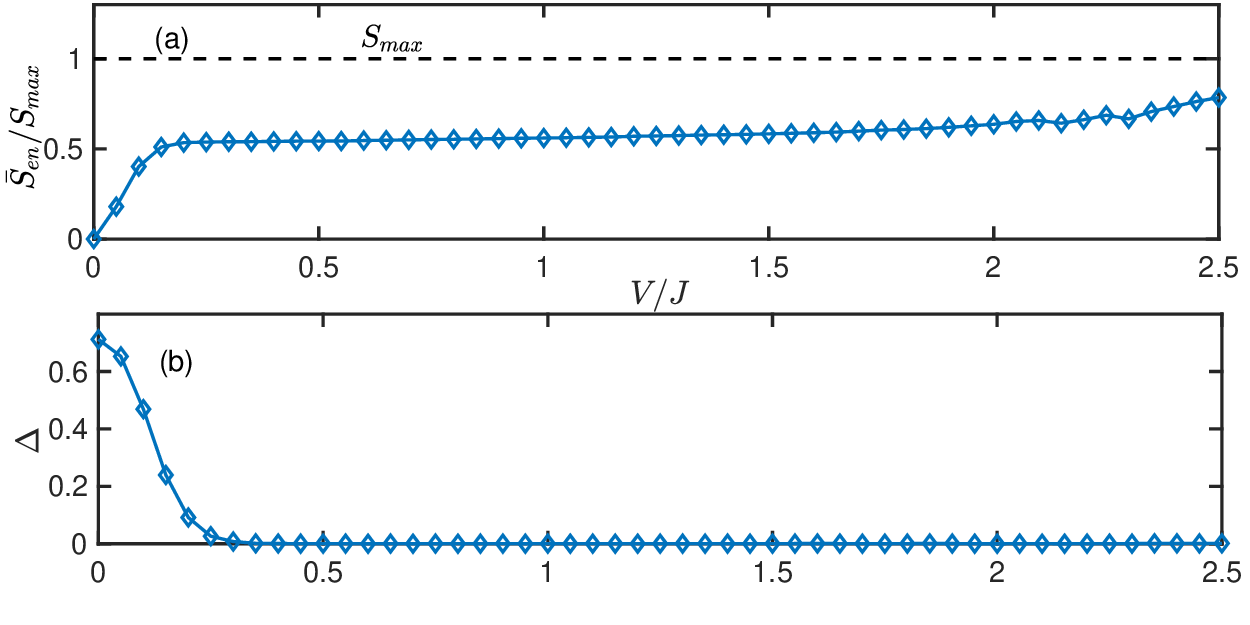}
\caption{Parameter dependence of long-time entanglement and energy convergence as functions of $V/J$. Panel (a) shows the ratio of the long-time averaged entanglement entropy to the maximum random-state entanglement entropy, where values near unity indicate strong entanglement. Panel (b) shows the normalized long-time averaged energy difference $\langle\Delta\rangle = \vert{}\langle E_A \rangle - \langle E_B \rangle\vert{}/(SJ)$, where values near zero indicate the emergence of MS. The initial state is chosen as $(z_1, \phi_1; z_2, \phi_2) = (0.01, 0; 0.8, 0)$. The long-time evolution is computed up to $T = 100/J$ with $S=20$ and $U/J=1$.}
\label{Fig5_0}
\end{figure}

Comparing panels (a) and (b) of Fig.~\ref{Fig5_0}, we observe a clear correspondence as the inter-species interaction $V/J$ increases. Specifically, the energy difference $\langle\Delta\rangle$ gradually decays toward zero, marking the onset of the MS state, while the normalized entanglement ratio $\langle S_{\mathrm{en}}\rangle/S_{\max}$ exhibits a pronounced monotonic increase toward saturation. This parameter correspondence reveals an intrinsic connection between quantum entanglement and macroscopic measure synchronization. 

Consistent with eigenstate thermalization, the progressive buildup of near-maximal entanglement provides an essential microscopic bridge that enables subsystems $A$ and $B$ to exchange information and equilibrate their energies. As analyzed in Fig.~\ref{Fig3}, the rate of entanglement growth varies across the integrable, near-chaotic, and deeply chaotic regimes and influences the temporal characteristics of synchronization. At the selected deeply chaotic point, rapid entanglement saturation allows MS to be established efficiently and with weak dependence on the initial configuration. Thus, while near-maximal entanglement originates from the underlying chaotic spectral properties, it provides a key microscopic ingredient that accompanies and facilitates the onset and robustness of macroscopic MS at the parameter points considered here.

\section{Conclusion}\label{V}
We have systematically investigated the realization and properties of MS in a two-component ultracold bosonic system, using a two-component bosonic Josephson junction as a concrete model. By comparatively analyzing the spectral statistics, subsystem energy expectation values, quantum fluctuations, and bipartite entanglement entropy, we have elucidated the microscopic mechanisms that accompany and facilitate macroscopic synchronization in the chaotic regime.

Our results show that the dynamical characteristics of MS change progressively as the system transitions from the integrable regime to the near-chaotic and deeply chaotic regimes. In the integrable and near-chaotic regimes, constrained phase-space structures and slow transient oscillations hinder or delay the establishment of MS and make it strongly sensitive to the initial conditions. By contrast, at the selected deeply chaotic parameter point, consistent with eigenstate thermalization and rapid near-maximal entanglement generation, MS is established with high efficiency, weak dependence on the initial configuration, and a characteristic anti-phase locking of quantum fluctuations. Long-time analyses further show that the buildup of strong quantum correlations provides a key microscopic bridge that accompanies energy equilibration and phase-space mixing. The fast MS with weak dependence on the initial state suggests possible applications in quantum metrology and sensing, as well as in quantum-state engineering in quantum simulators, for quantum many-body systems.

\section*{Acknowledgments}\label{VII}
This work was supported by the
Quantum Science and Technology–National Science and
Technology Major Project (No. 2021ZD0303200); the
National Natural Science Foundation of China
(No. 12374328, No. 11974116, and No. 12234014); the
Science and Technology Innovation Plan of Shanghai
Science and Technology Commission (No. 24LZ1400600);
the Shanghai Municipal Science and Technology
Major Project (No. 2019SHZDZX01); 
the National Key Research and Development Program of China
(No. 2016YFA0302001); the National Natural Science Foundation of China (Grant No.~12504566); the Natural Science Foundation of the Jiangsu Higher Education Institutions (Grant No.~25KJB140013); the Basic Research Program of Jiangsu (Grant No.~BK20250947); the Natural Science Foundation of Nantong City (Grant No.~JC2024045);
the Fundamental Research Funds
for the Central Universities; the Chinese National Youth
Talent Support Program, and the Shanghai Talent program.

\appendix
\section{Derivation Details of the Effective Spin Hamiltonian} \label{AA}

To derive the effective spin Hamiltonian in Eq.~(\ref{e3}) from Eq.~(\ref{e1}) in the main text, we first introduce the Schwinger-boson representation and define the spin operators within this representation. Because no inter-species exchange interaction is present, the particle number of each component is conserved separately. We therefore take the total particle number of each component to be $N$, i.e., $N_A=N_B=N$, and define the corresponding total spin quantum number as $S=N/2$. In this representation, the spin operators can be written as quadratic forms of creation and annihilation operators as follows:
\begin{equation}
\label{A1}
\hat{S}_{ix} = \frac{1}{2}(\hat{a}_{iR}^\dagger \hat{a}_{iL} + \hat{a}_{iL}^\dagger \hat{a}_{iR}), \quad \hat{S}_{iz} = \frac{1}{2}(\hat{n}_{iL} - \hat{n}_{iR}). 
\end{equation}
Here, $\hat{n}_{i\alpha} = \hat{a}_{i\alpha}^\dagger \hat{a}_{i\alpha}$ is the number operator. Since the total boson number operator $\hat{N}_i \equiv \hat{n}_{iL} + \hat{n}_{iR} $ commutes with the Hamiltonian and is conserved, we restrict our analysis to the subspace with a fixed particle number $\langle\hat{N}\rangle=N$. Within this subspace, the relation $\hat{n}_{iL} + \hat{n}_{iR} = N = 2S$ holds for each species $i$. From the definition of $\hat{S}_{iz}$ and the total number conservation, we can invert these relations to express the number operators for the left and right states in terms of the spin component
\begin{equation}
\label{A2}
\hat{n}_{iL} = S + \hat{S}_{iz}, \quad \hat{n}_{iR} = S - \hat{S}_{iz}.
\end{equation}
These two relations form the basis for all subsequent transformations, as they explicitly link the number operators to the spin operators. 

We now consider the interaction terms in the Hamiltonian. The first is the intra-well interaction term, i.e., the interaction between particles of the same component in the same well, whose original form is $\frac{U}{2N}\sum_{\alpha=L,R}\hat{n}_{i\alpha}(\hat{n}_{i\alpha}-1)$. Substituting the expressions for $\hat{n}_{iL}$ and $\hat{n}_{iR}$, we compute $\hat{n}_{iL}(\hat{n}_{iL}-1)$ and $\hat{n}_{iR}(\hat{n}_{iR}-1)$ separately and then sum them. In the expansion, we need the sum of squares:
\begin{equation}
\label{A3}
\hat{n}_{iL}^2 + \hat{n}_{iR}^2 = (S + \hat{S}_{iz})^2 + (S - \hat{S}_{iz})^2 = 2S^2 + 2\hat{S}_{iz}^2.
\end{equation}
Meanwhile, the sum of the linear terms is $\hat{n}_{iL} + \hat{n}_{iR} = 2S$. Therefore
\begin{equation}
\label{A4}
\sum_{\alpha} \hat{n}_{i\alpha}(\hat{n}_{i\alpha} - 1) = 2S(S-1) + 2\hat{S}_{iz}^2.
\end{equation}
Multiplying this result by the coefficient $\frac{U}{2N} = \frac{U}{4S}$, we obtain
\begin{equation}
\label{A5}
\frac{U}{2N} \sum_{\alpha} \hat{n}_{i\alpha}(\hat{n}_{i\alpha} - 1) = \frac{U}{2S}[S(S-1) + \hat{S}_{iz}^2].
\end{equation}
In the thermodynamic limit $(S \gg 1)$, we have $S(S-1) \approx S^2$. Neglecting the constant term $\frac{U}{2S} S^2$ which corresponds to the zero-point energy, the expression can finally be written as
\begin{equation}
\label{A6}
\frac{U}{2N} \sum_{\alpha} \hat{n}_{i\alpha}(\hat{n}_{i\alpha} - 1) = \frac{U}{4N}(N^2 + 4\hat{S}_{iz}^2).
\end{equation}
Here we have used $N = 2S,$ i.e., $N^2 = 4S^2$. 

Next, we consider the inter-species interaction term, which describes the interaction between particles of different components in the same well. Its original form is $\frac{V}{N}(\hat{n}_{AL}\hat{n}_{BL}+\hat{n}_{AR}\hat{n}_{BR})$. Substituting $\hat{n}_{iL} = S + \hat{S}_{iz}$ and $\hat{n}_{iR} = S - \hat{S}_{iz}$, we compute the two products separately
\begin{equation}
\begin{aligned}
\label{A7}
&\hat{n}_{AL}\hat{n}_{BL} = S^2 + S(\hat{S}_{Az} + \hat{S}_{Bz}) + \hat{S}_{Az}\hat{S}_{Bz},\\
&\hat{n}_{AR}\hat{n}_{BR} = S^2 - S(\hat{S}_{Az} + \hat{S}_{Bz}) + \hat{S}_{Az}\hat{S}_{Bz}.
\end{aligned}
\end{equation}
Adding the two expressions, the cross terms $\pm S(\hat{S}_{Az} + \hat{S}_{Bz})$ cancel each other, yielding
\begin{equation}
\label{A8}
\hat{n}_{AL}\hat{n}_{BL} + \hat{n}_{AR}\hat{n}_{BR} = 2S^2 + 2\hat{S}_{Az}\hat{S}_{Bz}.
\end{equation}
Multiplying by the coefficient $\frac{V}{N} = \frac{V}{2S}$, we get
\begin{equation}
\begin{aligned}
\label{A9}
\frac{V}{N}(\hat{n}_{AL}\hat{n}_{BL} + \hat{n}_{AR}\hat{n}_{BR}) = VS + \frac{V}{S}\hat{S}_{Az}\hat{S}_{Bz}.
\end{aligned}
\end{equation}
Again, neglecting the constant term $VS$ (which is part of the zero-point energy) and using $N = 2S$, we can rewrite the above as
\begin{equation}
\label{A10}
\frac{V}{N}(\hat{n}_{AL}\hat{n}_{BL} + \hat{n}_{AR}\hat{n}_{BR}) = \frac{V}{2N}(N^2 + 4\hat{S}_{Az}\hat{S}_{Bz}). 
\end{equation}

Finally, we include the tunneling term, i.e., the single-particle tunneling between the two wells. In the Schwinger-boson representation, this term becomes $-J(\hat{S}_{Ax}+\hat{S}_{Bx})$, where $J$ denotes the single-particle tunneling amplitude. Combining all contributions and neglecting constant zero-point energy terms, we obtain the effective Hamiltonian of the two-component bosonic Josephson junction
\begin{equation}
\label{A11}
\hat{H}=-J\hat{S}_{Ax}-J\hat{S}_{Bx}+\frac{U}{2S}\left(\hat{S}_{Az}^2+\hat{S}_{Bz}^2\right)+\frac{V}{S}\hat{S}_{Az}\hat{S}_{Bz}.
\end{equation}
Thus, applying the Schwinger-boson representation to Eq.~(\ref{e1}) and neglecting constant zero-point energy terms, we obtain Eq.~(\ref{e3}) in the main text.

\section*{References}
\bibliography{Ref_MS_Choas}

@book{Huygens1673,
    author = {Huygens, Christiaan},
    title = {Horologium Oscillatorium},
    publisher = {Apud F. Muquet},
    address = {Parisiis},
    year = {1673}
}

@article{Blasius1999,
    author = {Blasius, B. and Huppert, A. and Stone, L.},
    title = {Complex dynamics and phase synchronization in spatially extended ecological systems},
    journal = {Nature},
    volume = {399},
    number = {6734},
    pages = {354},
    year = {1999},
    month = {May},
    doi = {10.1038/20676},
    url = {https://doi.org/10.1038/20676}
}

@article{PoratShliom2014,
    author = {Porat-Shliom, N. and Chen, Y. and Tora, M. and Shitara, A. and Masedunskas, A. and Weigert, R.},
    title = {In vivo tissue-wide synchronization of mitochondrial metabolic oscillations},
    journal = {Cell Reports},
    volume = {9},
    number = {2},
    pages = {514--521},
    year = {2014},
    month = {oct},
    doi = {10.1016/j.celrep.2014.09.022},
    url = {https://doi.org/10.1016/j.celrep.2014.09.022}
}

@article{Zhirov2009,
    author = {Zhirov, O. V. and Shepelyansky, D. L.},
    title = {Quantum synchronization and entanglement of two qubits coupled to a driven dissipative resonator},
    journal = {Phys. Rev. B},
    volume = {80},
    pages = {014519},
    year = {2009},
    doi = {10.1103/PhysRevB.80.014519},
    url = {https://doi.org/10.1103/PhysRevB.80.014519}
}

@article{Orth2010,
    author = {Orth, P. P. and Roosen, D. and Hofstetter, W. and LeHur, K.},
    title = {Dynamics, synchronization, and quantum phase transitions of two dissipative spins},
    journal = {Phys. Rev. B},
    volume = {82},
    pages = {144423},
    year = {2010},
    doi = {10.1103/PhysRevB.82.144423},
    url = {https://doi.org/10.1103/PhysRevB.82.144423}
}

@article{Hampton1999,
    author = {Hampton, Alan and Zanette, Dami{\'a}n H.},
    title = {Measure Synchronization in Coupled Hamiltonian Systems},
    journal = {Phys. Rev. Lett.},
    volume = {83},
    number = {11},
    pages = {2179--2182},
    year = {1999},
    doi = {10.1103/PhysRevLett.83.2179},
    url = {https://doi.org/10.1103/PhysRevLett.83.2179}
}

@article{Ghosh2023,
    author = {Ghosh, Anupam},
    title = {Measure synchronization in interacting Hamiltonian systems: A brief review},
    journal = {Chaos Solitons \& Fractals},
    volume = {177},
    pages = {114237},
    year = {2023},
    doi = {10.1016/j.chaos.2023.114237},
    url = {https://doi.org/10.1016/j.chaos.2023.114237}
}

@article{Mari2013,
    author = {Mari, A. and Farace, A. and Didier, N. and Giovannetti, V. and Fazio, R.},
    title = {Measures of Quantum Synchronization in Continuous Variable Systems},
    journal = {Phys. Rev. Lett.},
    volume = {111},
    number = {10},
    pages = {103605},
    year = {2013},
    doi = {10.1103/PhysRevLett.111.103605},
    url = {https://doi.org/10.1103/PhysRevLett.111.103605}
}

@article{Wang2003,
    author = {Wang, Xingang and Zhan, Meng and Lai, C.-H. and Gang, Hu},
    title = {Measure synchronization in coupled Hamiltonian systems},
    journal = {Phys. Rev. E},
    volume = {68},
    number = {6},
    pages = {066208},
    year = {2003},
    doi = {10.1103/PhysRevE.68.066208},
    url = {https://doi.org/10.1103/PhysRevE.68.066208}
}

@article{Bemani2017,
    author = {Bemani, F. and Motazedifard, A. and Roknizadeh, R. and Naderi, M. H. and Vitali, D.},
    title = {Synchronization dynamics of two nanomechanical membranes within a Fabry-Perot cavity},
    journal = {Phys. Rev. A},
    volume = {96},
    number = {2},
    pages = {023805},
    year = {2017},
    month = {Aug},
    doi = {10.1103/PhysRevA.96.023805},
    url = {https://doi.org/10.1103/PhysRevA.96.023805}
}

@book{Pikovsky2001,
    author = {Pikovsky, Arkady and Rosenblum, Michael and Kurths, J{\"u}rgen},
    title = {Synchronization: {A} Universal Concept in Nonlinear Sciences},
    publisher = {Cambridge University Press},
    address = {New York},
    year = {2001},
    doi = {10.1017/CBO9780511755743},
    url = {https://doi.org/10.1017/CBO9780511755743},
    isbn = {978-0-521-53352-2}
}

@article{Qiu2014,
    author = {Qiu, Haibo and Juli{\'a}-D{\'i}az, Bruno and Garcia-March, Miguel Angel and Polls, Artur},
    title = {Measure synchronization in quantum many-body systems},
    journal = {Phys. Rev. A},
    volume = {90},
    number = {3},
    pages = {033603},
    year = {2014},
    doi = {10.1103/PhysRevA.90.033603},
    url = {https://doi.org/10.1103/PhysRevA.90.033603}
}

@article{Sur2020,
    author = {Sur, Saikat and Ghosh, Anupam},
    title = {Quantum counterpart of Measure synchronization: A study on a pair of Harper systems},
    journal = {Phys. Lett. A},
    volume = {384},
    number = {8},
    pages = {126176},
    year = {2020},
    doi = {10.1016/j.physleta.2019.126176},
    url = {https://doi.org/10.1016/j.physleta.2019.126176}
}

@article{Oganesyan2007,
    author = {Oganesyan, Vadim and Huse, David A.},
    title = {Localization of interacting fermions at high temperature},
    journal = {Phys. Rev. B},
    volume = {75},
    number = {15},
    pages = {155111},
    year = {2007},
    doi = {10.1103/PhysRevB.75.155111},
    url = {https://doi.org/10.1103/PhysRevB.75.155111}
}

@article{Atas2013,
    author = {Atas, Y. Y. and Bogomolny, E. and Giraud, O. and Roux, G.},
    title = {Distribution of the ratio of consecutive level spacings in random matrix ensembles},
    journal = {Phys. Rev. Lett.},
    volume = {110},
    number = {8},
    pages = {084101},
    year = {2013},
    doi = {10.1103/PhysRevLett.110.084101},
    url = {https://doi.org/10.1103/PhysRevLett.110.084101}
}

@book{Haake2010,
    author = {Haake, Fritz},
    title = {Quantum Signatures of Chaos},
    publisher = {Springer},
    address = {Berlin},
    year = {2010},
    edition = {3rd},
    doi = {10.1007/978-3-642-05428-0},
    url = {https://doi.org/10.1007/978-3-642-05428-0},
    isbn = {978-3-642-05427-3}
}

@article{Leggett2001,
    author = {Leggett, Anthony J.},
    title = {Bose-Einstein condensation in the alkali gases: Some fundamental concepts},
    journal = {Rev. Mod. Phys.},
    volume = {73},
    number = {2},
    pages = {307--356},
    year = {2001},
    month = {April},
    doi = {10.1103/RevModPhys.73.307},
    url = {https://doi.org/10.1103/RevModPhys.73.307}
}

@article{Robb1998,
    author = {Robb, Daniel T. and Reichl, L. E.},
    title = {Chaos in a two-spin system with applied magnetic field},
    journal = {Phys. Rev. E},
    volume = {57},
    number = {2},
    pages = {2458--2459},
    year = {1998},
    month = {February},
    doi = {10.1103/PhysRevE.57.2458},
    url = {https://doi.org/10.1103/PhysRevE.57.2458},
    publisher = {American Physical Society}
}

@article{Emerson2001,
    author = {Emerson, J. and Ballentine, L. E.},
    title = {Characteristics of quantum-classical correspondence for two interacting spins},
    journal = {Phys. Rev. A},
    volume = {63},
    number = {5},
    pages = {052103},
    year = {2001},
    month = {April},
    doi = {10.1103/PhysRevA.63.052103},
    url = {https://doi.org/10.1103/PhysRevA.63.052103},
    publisher = {American Physical Society}
}

@article{Milburn1997,
    author = {Milburn, G. J. and Corney, J. and Wright, E. M. and Walls, D. F.},
    title = {Quantum dynamics of an atomic Bose-Einstein condensate in a double-well potential},
    journal = {Phys. Rev. A},
    volume = {55},
    number = {6},
    pages = {4318--4324},
    year = {1997},
    month = {June},
    doi = {10.1103/PhysRevA.55.4318},
    url = {https://doi.org/10.1103/PhysRevA.55.4318}
}

@article{Wang2003b,
    author = {Wang, Xingang and Zhan, Meng and Lai, C.-H. and Gang, Hu},
    title = {Measure synchronization in coupled {\(\varphi$}$^{4}$ {H}amiltonian systems},
    journal = {Phys. Rev. E},
    volume = {67},
    number = {6},
    pages = {066215},
    year = {2003},
    month = {jun},
    doi = {10.1103/PhysRevE.67.066215},
    url = {https://doi.org/10.1103/PhysRevE.67.066215},
    pmid = {16241334}
}

@article{Page1993,
    author = {Page, Don N.},
    title = {Average Entropy of a Subsystem},
    journal = {Phys. Rev. Lett.},
    volume = {71},
    number = {9},
    pages = {1291--1294},
    year = {1993},
    doi = {10.1103/PhysRevLett.71.1291},
    url = {https://doi.org/10.1103/PhysRevLett.71.1291},
    archivePrefix = {arXiv},
    eprint = {gr-qc/9305007},
    primaryClass = {gr-qc}
}

@article{Radcliffe1971,
    author = {Radcliffe, J. M.},
    title = {Some properties of coherent spin states},
    journal = {J. Phys. A: Gen. Phys.},
    volume = {4},
    number = {3},
    pages = {313--323},
    year = {1971},
    doi = {10.1088/0305-4470/4/3/009},
    url = {https://doi.org/10.1088/0305-4470/4/3/009}
}

@article{Berry1977,
    author = {Berry, M. V. and Tabor, M.},
    title = {Level clustering in the regular spectrum},
    journal = {Proc. R. Soc. Lond. A},
    volume = {356},
    number = {1686},
    pages = {375--394},
    year = {1977},
    doi = {10.1098/rspa.1977.0140},
    url = {https://doi.org/10.1098/rspa.1977.0140}
}

@article{Aguiar2011,
    author = {Aguiar, Manuela and Ashwin, Peter and Dias, Ana and Field, Michael},
    title = {Dynamics of coupled cell networks: Synchrony, heteroclinic cycles and inflation},
    journal = {J. Nonlin. Sci.},
    volume = {21},
    number = {2},
    pages = {271--323},
    year = {2011},
    doi = {10.1007/s00332-010-9080-1},
    url = {https://doi.org/10.1007/s00332-010-9080-1}
}

@book{Kuramoto1984,
    author = {Kuramoto, Yoshiki},
    title = {Chemical Oscillations, Waves and Turbulence},
    publisher = {Springer},
    address = {Berlin},
    year = {1984},
    doi = {10.1007/978-3-642-69689-3},
    url = {https://doi.org/10.1007/978-3-642-69689-3},
    isbn = {978-3-642-69691-6}
}

@article{Arenas2008,
    author = {Arenas, Alex and D{\'i}az-Guilera, Albert and Kurths, J{\"u}rgen and Moreno, Yamir and Zhou, Changsong},
    title = {Synchronization in complex networks},
    journal = {Phys. Rep.},
    volume = {469},
    number = {3},
    pages = {93--153},
    year = {2008},
    doi = {10.1016/j.physrep.2008.09.002},
    url = {https://doi.org/10.1016/j.physrep.2008.09.002}
}

@article{Manzano2013,
  author = {G. Manzano and F. Galve and G. L. Giorgi and E. Hern\'andez-Garc\'{\i}a and R. Zambrini},
  title = {Synchronization, quantum correlations and entanglement in oscillator networks},
  journal = {Sci. Rep.},
  volume = {3},
  pages = {1439},
  year = {2013},
  doi = {10.1038/srep01439},
  url = {https://doi.org/10.1038/srep01439}
}

@article{Horodecki2009,
  author = {R. Horodecki and P. Horodecki and M. Horodecki and K. Horodecki},
  title = {Quantum entanglement},
  journal = {Rev. Mod. Phys.},
  volume = {81},
  pages = {865},
  year = {2009},
  doi = {10.1103/RevModPhys.81.865},
  url = {https://doi.org/10.1103/RevModPhys.81.865}
}

@article{Neill2016,
  author = {C. Neill and P. Roushan and M. Fang and Y. Chen and M. Kolodrubetz and Z. Chen and A. Megrant and R. Barends and B. Campbell and B. Chiaro and others},
  title = {Ergodic dynamics and thermalization in an isolated quantum system},
  journal = {Nat. Phys.},
  volume = {12},
  pages = {1037},
  year = {2016},
  doi = {10.1038/nphys3830},
  url = {https://doi.org/10.1038/nphys3830}
}

@article{Piga2019,
  author = {A. Piga and M. Lewenstein and J. Q. Quach},
  title = {Quantum chaos and entanglement in ergodic and nonergodic systems},
  journal = {Phys. Rev. E},
  volume = {99},
  pages = {032213},
  year = {2019},
  doi = {10.1103/PhysRevE.99.032213},
  url = {https://doi.org/10.1103/PhysRevE.99.032213}
}

@article{Rosa2020,
  author = {D. Rosa and D. Rossini and G. Andolina and M. Polini and M. Carrega},
  title = {Ultra-stable charging of fast-scrambling SYK quantum batteries},
  journal = {J. High Energy Phys.},
  volume = {11},
  pages = {067},
  year = {2020},
  doi = {10.1007/JHEP11(2020)067},
  url = {https://doi.org/10.1007/JHEP11(2020)067}
}

@article{Schack1994,
  author = {R. Schack and G. M. D'Ariano and C. M. Caves},
  title = {Hypersensitivity to perturbation in the quantum kicked top},
  journal = {Phys. Rev. E},
  volume = {50},
  pages = {972},
  year = {1994},
  doi = {10.1103/PhysRevE.50.972},
  url = {https://doi.org/10.1103/PhysRevE.50.972}
}

@article{Ghose2008,
  author = {S. Ghose and R. Stock and P. Jessen and R. Lal and A. Silberfarb},
  title = {Chaos, entanglement, and decoherence in the quantum kicked top},
  journal = {Phys. Rev. A},
  volume = {78},
  pages = {042318},
  year = {2008},
  doi = {10.1103/PhysRevA.78.042318},
  url = {https://doi.org/10.1103/PhysRevA.78.042318}
}

@article{Lee2001,
  author = {C. Lee and W. Hai and L. Shi and X. Zhu and K. Gao},
  title = {Chaotic and frequency-locked atomic population oscillations between two coupled Bose-Einstein condensates},
  journal = {Phys. Rev. A},
  volume = {64},
  pages = {053604},
  year = {2001},
  doi = {10.1103/PhysRevA.64.053604},
  url = {https://doi.org/10.1103/PhysRevA.64.053604}
}

@article{Milonni1983,
  author = {P. W. Milonni and J. R. Ackerhalt and H. W. Galbraith},
  title = {Chaos in the Semiclassical N-Atom Jaynes-Cummings Model: Failure of the Rotating-Wave Approximation},
  journal = {Phys. Rev. Lett.},
  volume = {50},
  pages = {966},
  year = {1983},
  doi = {10.1103/PhysRevLett.50.966},
  url = {https://doi.org/10.1103/PhysRevLett.50.966}
}

@article{Furuya1998,
  author = {K. Furuya and M. C. Nemes and G. Q. Pellegrino},
  title = {Quantum Dynamical Manifestation of Chaotic Behavior in the Process of Entanglement},
  journal = {Phys. Rev. Lett.},
  volume = {80},
  pages = {5524},
  year = {1998},
  doi = {10.1103/PhysRevLett.80.5524},
  url = {https://doi.org/10.1103/PhysRevLett.80.5524}
}

@article{Emary2003,
  author = {C. Emary and T. Brandes},
  title = {Chaos and the quantum phase transition in the Dicke model},
  journal = {Phys. Rev. E},
  volume = {67},
  pages = {066203},
  year = {2003},
  doi = {10.1103/PhysRevE.67.066203},
  url = {https://doi.org/10.1103/PhysRevE.67.066203}
}

@article{Kolovsky2004,
  author = {A. R. Kolovsky and A. Buchleitner},
  title = {Quantum chaos in the Bose-Hubbard model},
  journal = {Europhys. Lett.},
  volume = {68},
  pages = {632},
  year = {2004},
  doi = {10.1209/epl/i2004-10258-8},
  url = {https://doi.org/10.1209/epl/i2004-10258-8}
}

@article{Kolovsky2016,
  author = {A. R. Kolovsky},
  title = {Bose-Hubbard Hamiltonian: Quantum chaos approach},
  journal = {Int. J. Mod. Phys. B},
  volume = {30},
  pages = {1630009},
  year = {2016},
  doi = {10.1142/S0217979216300096},
  url = {https://doi.org/10.1142/S0217979216300096}
}

@article{Giraldo2020,
  author = {A. Giraldo and B. Krauskopf and N. G. R. Broderick and J. A. Levenson and A. M. Yacomotti},
  title = {The driven-dissipative Bose-Hubbard dimer: Phase diagram and chaos},
  journal = {New J. Phys.},
  volume = {22},
  pages = {043009},
  year = {2020},
  doi = {10.1088/1367-2630/ab7a82},
  url = {https://doi.org/10.1088/1367-2630/ab7a82}
}

@incollection{Casati1979,
  author = {G. Casati and B. V. Chirikov and F. M. Izrailev and J. Ford},
  title = {Stochastic Behavior in Classical and Quantum Hamiltonian Systems},
  booktitle = {Stochastic Behavior in Classical and Quantum Hamiltonian Systems},
  editor = {G. Casati and J. Ford},
  publisher = {Springer},
  address = {Berlin},
  pages = {334--352},
  year = {1979},
  doi = {10.1007/BFb0013485},
  url = {https://doi.org/10.1007/BFb0013485}
}

@article{Eckmann1985,
  author = {J.-P. Eckmann and D. Ruelle},
  title = {Ergodic theory of chaos and strange attractors},
  journal = {Rev. Mod. Phys.},
  volume = {57},
  pages = {617},
  year = {1985},
  doi = {10.1103/RevModPhys.57.617},
  url = {https://doi.org/10.1103/RevModPhys.57.617}
}

@article{Bohigas1984a,
  author = {O. Bohigas and M. J. Giannoni and C. Schmit},
  title = {Characterization of Chaotic Quantum Spectra and Universality of Level Fluctuation Laws},
  journal = {Phys. Rev. Lett.},
  volume = {52},
  pages = {1},
  year = {1984},
  doi = {10.1103/PhysRevLett.52.1},
  url = {https://doi.org/10.1103/PhysRevLett.52.1}
}

@article{Bohigas1984b,
  author = {O. Bohigas and M. J. Giannoni and C. Schmit},
  title = {Spectral fluctuations of classically chaotic quantum systems},
  journal = {J. Phys. Lett.},
  volume = {45},
  pages = {1015},
  year = {1984},
  doi = {10.1051/jphyslet:0198400450210101500},
  url = {https://doi.org/10.1051/jphyslet:0198400450210101500}
}

@article{Eckmann1985Ergodic,
  title = {Ergodic theory of chaos and strange attractors},
  author = {Eckmann, J.-P. and Ruelle, D.},
  journal = {Rev. Mod. Phys.},
  volume = {57},
  number = {3},
  pages = {617--656},
  year = {1985},
  publisher = {American Physical Society},
  doi = {10.1103/RevModPhys.57.617},
  note = {Addendum: Rev. Mod. Phys. 57, 1115 (1985)}
}

@article{Mondal2020Chaos,
  title = {Chaos and quantum scars in a coupled top model},
  author = {Mondal, Debabrata and Sinha, Sudip and Sinha, S.},
  journal = {Phys. Rev. E},
  volume = {102},
  issue = {2},
  pages = {020101(R)},
  year = {2020},
  month = {Aug},
  publisher = {American Physical Society},
  doi = {10.1103/PhysRevE.102.020101},
  url = {https://link.aps.org/doi/10.1103/PhysRevE.102.020101}
}

@article{PhysRevLett.111.103605,
  title = {Measures of Quantum Synchronization in Continuous Variable Systems},
  author = {Mari, A. and Farace, A. and Didier, N. and Giovannetti, V. and Fazio, R.},
  journal = {Phys. Rev. Lett.},
  volume = {111},
  issue = {10},
  pages = {103605},
  numpages = {5},
  year = {2013},
  month = {Sep},
  publisher = {American Physical Society},
  doi = {10.1103/PhysRevLett.111.103605},
  url = {https://link.aps.org/doi/10.1103/PhysRevLett.111.103605}
}

\end{document}